\documentclass[twocolumn,aps,prl,superscriptaddress,nofootinbib]{revtex4-1}
\usepackage{graphicx}
\graphicspath{{figures/}{fig/}}
\DeclareGraphicsExtensions{.eps}
\usepackage{amsmath}
\usepackage{bm}
\usepackage{color}
\usepackage{slashed}
\usepackage{epsfig}
\usepackage{amsfonts}
\usepackage{epstopdf}
\usepackage[pdftex]{hyperref}
\usepackage{extarrows}
\usepackage{mathrsfs}
\newcommand{\ben}{\begin{eqnarray}}
\newcommand{\een}{\end{eqnarray}}

\newcommand{\bef}{\begin{figure}[!htp]}
\newcommand{\eef}{\end{figure}}

\newcommand{\bea}{\begin{eqnarray}}
\newcommand{\eea}{\end{eqnarray}}

\def\ba{\begin{linenomath*}\begin{equation}}
\def\ea{\end{equation}\end{linenomath*}}

\allowdisplaybreaks

\newcommand{\sect}[1]{\section{#1}}
\begin{document}
\title{Extraction of Next-to-Next-to-Leading-Order PDFs from Lattice QCD Calculations}

\author{Zheng-Yang Li}
\email{lizhengyang@pku.edu.cn}
\affiliation{School of Physics and State Key Laboratory of Nuclear Physics and
	Technology, Peking University, Beijing 100871, China}
\author{Yan-Qing Ma}
\email{yqma@pku.edu.cn}
\affiliation{School of Physics and State Key Laboratory of Nuclear Physics and
	Technology, Peking University, Beijing 100871, China}
\affiliation{Center for High Energy physics, Peking University, Beijing 100871, China}
\affiliation{Collaborative Innovation Center of Quantum Matter,
	Beijing 100871, China}
\author{Jian-Wei Qiu}
\email{jqiu@jlab.org}
\affiliation{Theory Center, Jefferson Lab, 12000 Jefferson Avenue, Newport News, VA 23606, USA}

\date{\today}

\begin{abstract}
{ 
We present for the first time complete next-to-next-to-leading-order coefficient functions to match flavor non-singlet 
quark correlation functions in position space, which are calculable in lattice QCD, to parton distribution functions (PDFs).  
Using PDFs extracted from experimental data and our calculated matching coefficients, 
we predict valence-quark correlation functions that can be confronted by lattice QCD calculations.
The uncertainty of our predictions is greatly reduced with higher order matching coefficients.} { By performing Fourier transformation, we also obtain matching coefficients for corresponding quasi-PDFs and pseudo-PDFs.}
Our method of calculations can be readily generalized to evaluate the matching coefficients for sea-quark and gluon correlation functions, putting the program to extract partonic structure of hadrons from lattice QCD calculations to be comparable with { and complementary to} that from experimental measurements.
\end{abstract}

\maketitle
\allowdisplaybreaks

\sect{Introduction}	
\label{sec:intro}
Parton distribution functions (PDFs) encode important nonperturbative information of strong interactions, and they are crucial for understanding all phenomena at the Large Hadron Colliders (LHC)~\cite{Lin:2020rut}. In terms of QCD factorization~\cite{Collins:1989gx}, a typical hadronic cross section with a large momentum transfer $Q$ and collision energy $\sqrt{S}$ at the LHC can be factorized as
\begin{align}\label{eq:fac-hh}
d\sigma_{hh'}(Q^2,S) &
= \sum_{i,j} f_{i/h}(x,\mu^2) \otimes f_{j/h'}(x',\mu^2) \nonumber
\\
& {\hskip -0.2in}
\otimes d\hat{\sigma}_{ij}(x,x',\mu^2, Q^2,S)
+{ O}(\Lambda_{\rm QCD}^2/Q^2) \,,
\end{align}
where $i,j=q,\bar{q},g$ represents parton flavor, $f_{i/h}(x,\mu^{{2}})$ is the PDF as a probability distribution to find an active parton of flavor $i$ inside a colliding hadron $h$ with the parton carrying the hadron's momentum fraction $x$, probed at a factorization scale $\mu\sim { O}(Q)$, $d\hat{\sigma}_{ij}$ represents a short-distance partonic scattering, and $\otimes$ indicates an integration over value of $x$ or $x'$, accessible by the scattering cross section.  By measuring hadronic cross sections, with perturbatively calculated partonic hard parts $d\hat{\sigma}_{ij}$, PDFs have been extracted from the world data at the state-of-the-art next-to-next-to-leading order (NNLO) accuracy~\cite{Lin:2020rut}.

With the steep falling nature of PDFs as $x\to 1$ and the convolution in Eq.~(\ref{eq:fac-hh}), the uncertainty of extracted PDFs at large $x$ is so significant that limits our confidence to push the search for signals of new physics to larger invariant mass.  With the nonperturbative nature of PDFs, it is natural to ask if we can calculate PDFs directly in lattice QCD (LQCD).  A short answer is no since the operators defining PDFs are time-dependent and LQCD is formulated in Euclidean space-time.  Recently, stimulated by the quasi-PDFs approach~\cite{Ji:2013dva} (it was later formulated in a large-momentum effective field theory \cite{Ji:2014gla,Ji:2020ect}),
extraction of PDFs from lattice QCD calculation has drawn a lot of attentions and many 
{ new}
ideas appeared, including the pseudo-PDFs~\cite{Radyushkin:2017cyf}, current-current correlators in momentum space  \cite{Chambers:2017dov} and current-current correlators in position space~\cite{Ma:2017pxb}. See also some earlier related approaches~\cite{Liu:1993cv,Liu:1999ak,Liu:2016djw,Aglietti:1998ur,Abada:2001if,Braun:2007wv}.

As proposed by two of us in Refs.~\cite{Ma:2014jla,Ma:2017pxb}, PDFs can be extracted from any {\em good} LQCD observables, which was referred to as ``Lattice Cross Sections" (LCSs), that are calculable in LQCD and factorizable into PDFs with perturbatively calculable matching coefficients,
\begin{align}\label{eq:lcs}
\sigma_{n/h}(\omega,\xi^2)
&\equiv \langle h(p)| T\{{\cal O}_n(\xi)\} |h(p)\rangle
\nonumber \\
&=\sum_i f_{i/h}(x,\mu^{2}) \otimes K_{n/i}(x\omega, \xi^2, \mu^2)
\\
& {\hskip 0.4in}
+{ O}(\xi^2\Lambda_{\rm QCD}^2)\, ,
\nonumber
\end{align}
where $\xi$ with $\xi^2\neq 0$ represents the size of nonlocal operator ${\cal O}_n(\xi)$ of type $n$, controlling the short-distance physics of the factorization, $\omega\equiv p\cdot\xi$ (often referred as Ioffe time),
{ and $K_{n/i}$ are perturbative matching coefficients.  
The PDFs in Eqs.~(\ref{eq:fac-hh}) and (\ref{eq:lcs}) are the same, and}
can be extracted by QCD global fits of data generated by LQCD calculation of $\sigma_{n/h}(\omega,\xi^2)$ with various operator type $n$, 
{
together with}
the world data on various high energy scattering cross sections~\cite{Lin:2020rut,Ma:2014jla,Ma:2017pxb,Bringewatt:2020ixn}.

{
One key difference between Eqs.~(\ref{eq:fac-hh}) and (\ref{eq:lcs}) is that}
$\sigma_{n/h}$ in Eq.~(\ref{eq:lcs}) 
{
is not an experimentally measured physical cross section.} { The corresponding operator ${\cal O}_n(\xi)$, }{ which
can be a two-quark correlation operator that defines}
quasi-PDFs~\cite{Ji:2013dva}, current-current correlators~\cite{Ma:2017pxb}, or any 
{
others}
that satisfy the aforementioned properties, 
might require additional ultraviolet (UV) renormalization beyond using renormalized fields.  
This additional UV renormalization 
{ impacts the}
calculation and stability of the 
{ perturbative}
matching coefficients $K_{n/i}$ for LQCD observables.
Although extraction of PDFs from LQCD calculations have made tremendous progresses in recent years \cite{Ishikawa:2016znu,Chen:2016fxx,Monahan:2016bvm,Briceno:2017cpo, Xiong:2017jtn, Li:2018tpe,Zhang:2018diq, Ji:2017oey,Ishikawa:2017faj,Green:2017xeu,Constantinou:2017sej,Alexandrou:2017huk,Chen:2017mzz,Stewart:2017tvs,Wang:2019tgg,Xiong:2013bka,Ji:2017rah,Radyushkin:2017lvu,Izubuchi:2018srq,Orginos:2017kos,Sufian:2020vzb,Lin:2014zya,Alexandrou:2015rja,Chen:2016utp,Alexandrou:2016jqi,Zhang:2017bzy,Monahan:2017hpu,Ishikawa:2019flg,Alexandrou:2018pbm,Chen:2018xof,Chen:2018fwa,Liu:2018uuj,Bali:2018spj,Radyushkin:2018nbf,Lin:2018pvv,Karpie:2018zaz,Joo:2019bzr,Chai:2020nxw,Lin:2020ssv}, the state-of-the-art calculation of short-distance matching coefficients is still limited to the next-to-leading order (NLO) in almost all existing approaches~\cite{Xiong:2013bka,Ji:2017rah,Radyushkin:2017lvu,Izubuchi:2018srq,Stewart:2017tvs,Wang:2019tgg,Orginos:2017kos,Sufian:2020vzb}, which is partially limited by this additional renormalization and our ability to do perturbative calculation in coordinate space.  In this Letter, we derive for the first time the NNLO non-singlet matching coefficients in dimensional regularization, allowing us to extract PDFs from LQCD calculations at the same rigor as those extracted from experimental data, 
{ as well as}
addressing concerns that 
{ the factorization might be invalidated}
at NNLO~\cite{Li:2016amo}.

\sect{Quark correlation functions}	
\label{sec:qcf}
We focus on the following unpolarized gauge invariant quark correlation operator~\cite{Ji:2013dva}
\begin{align}\label{eq:qqo}
\begin{split}
{\cal O}_{q}^{\nu,b}(\xi,\mu^2,\delta) = \overline{\psi}_q(\xi)\,\gamma^\nu \Phi^{(f)}(\{\xi,0\})\,\psi_q(0)\big|_{\mu^2,\delta}\,,
\end{split}
\end{align}
which is made of renormalized fields with a path ordered gauge link in the fundamental representation, $\Phi^{(f)}(\xi,0)={\cal P}e^{-ig_s\int_0^{1} \xi\cdot A^{(f)}(r \xi)\,dr}$. Because this composite quark correlation operator is UV divergent, a UV regulator $\delta$ is needed, which may represent lattice spacing $a$ in lattice QCD calculations, or represent $\epsilon \equiv (4-d)/2$ in dimensional regularization (DR) of continuum calculations. $\mu$ is a dimensional scale accompanied by the UV regulator, which is different from the factorization scale in Eq.~(\ref{eq:lcs}), while one could choose them to be equal numerically. This UV divergence is multiplicatively renormalizable~\cite{Ji:2017oey,Ishikawa:2017faj,Green:2017xeu}, as
\begin{align}\label{eq:rqqo}
\begin{split}
{\cal O}_{q}^{\nu,\text{RS}}(\xi) ={\cal O}_{q}^{\nu,b}(\xi,\mu^2,\delta)/Z^{\text{RS}}(\xi^2, \mu^2,\delta) \,,
\end{split}
\end{align}
where superscript $\text{RS}$ indicates a renormalization scheme and $Z^{\text{RS}}(\xi^2,\mu^2,\delta)$ is the multiplicative renormalization constant. For regularization-invariant renormalization conditions, the renormalized ${\cal O}_{q}^{\nu,\text{RS}}$ are independent of $\delta$ and $\mu^2$.

Quark correlation functions (QCFs) are defined as hadronic matrix elements of ${\cal O}_{q}^{\nu,\text{RS}}(\xi)$
\begin{align}\label{eq:lcsq}
F_{q/h}^{\nu,\text{RS}}(\omega,\xi^2)
&= \langle h(p)| {\cal O}_{q}^{\nu,\text{RS}}(\xi) | h(p) \rangle\,,
\end{align}
which is independent of regularization scheme and scale, like physical cross sections.
With $\xi^0=0$ and $\xi^2\Lambda_{\rm QCD}^2\ll 1$, $F_{q/h}^{\nu,\text{RS}}(\omega,\xi^2)$
are expected to be calculable in LQCD and 
{ proved to be}
factorizable into PDFs~\cite{Ma:2014jla,Ma:2017pxb,Izubuchi:2018srq}.
Their Fourier transform over $d\omega$ with fixed $p$ leads to the quasi-PDFs; and with fixed $\xi$ is proportional to pseudo-PDFs \cite{Ma:2017pxb}.
In this Letter, we focus on flavor non-singlet combinations of QCFs, and have corresponding factorization formula in continuum
as~\cite{Ma:2017pxb}
\begin{align}\label{eq:factorization}
F_{q_{ik}/h}^{\nu,\text{RS}}(\omega,\xi^2)
&= \frac{1}{R^{\text{RS}}(\xi^2,\mu^2)} \int_{-1}^1 \frac{dx}{x} \, f_{q_{ik}/h}(x,\mu^2)
\\
& {\hskip 0.4in}\times K^\nu(x\omega,\xi^2, \mu^2)
+{ O}(\xi^2\Lambda_{\rm QCD}^2)\, ,
\nonumber
\end{align}
where  $R^{\text{RS}}(\xi^2,\mu^2)\equiv Z^{\text{RS}}(\xi^2,\mu^2,\epsilon)/Z^{\overline{\text{MS}}}(\xi^2,\mu^2,\epsilon)$ is a {\it finite} renormalization factor that transforms any ``preferred'' regularization-invariant $\text{RS}$ scheme to the conventional $\overline{\text{MS}}$ scheme, $K^\nu$ are perturbative matching coefficients in $\overline{\text{MS}}$ scheme, and $q_{ik} \equiv q_i-q_k$ 
means
\begin{align}
f_{q_{ik}/h}(x,\mu^2)&\equiv f_{q_i/h}(x,\mu^2)-f_{q_k/h}(x,\mu^2)\,,\\
F_{q_{ik}/h}^{\nu,\text{RS}}(\omega,\xi^2)&\equiv F_{q_i/h}^{\nu,\text{RS}}(\omega,\xi^2)-F_{q_k/h}^{\nu,\text{RS}}(\omega,\xi^2) \,,
\end{align}
where $q_i, q_k=u, d, s$ are quark flavors.
To extract the non-singlet distribution $f_{q_{ik}/h}$ from LQCD calculations of $F_{q_{ik}/h}^{\nu,\text{RS}}$ to the NNLO accuracy, we have to perturbatively calculate $R^{\text{RS}}$ and $K^\nu$ to the power of $\alpha_s^2$.
{ 
The factorization formula in Eq.~\eqref{eq:factorization} is also valid for valence-quark correlation functions by}
replacing $q_{ij}$ and $K^\nu(x\omega,\xi^2, \mu^2)$ 
{ with} 
$q_{v}\equiv q-\bar{q}$ and $K_v^\nu(x\omega,\xi^2, \mu^2)\equiv K^\nu(x\omega,\xi^2, \mu^2)-K^\nu(-x\omega,\xi^2, \mu^2)$, respectively.

\sect{Renormalization constant}
\label{sec:vev}
The renormalization constant $Z^{\rm RS}$ introduced in Eq.~(\ref{eq:rqqo}) is determined by 
short-distance property of the quark correlation operator in Eq.~(\ref{eq:qqo}) and should not depend on the hadronic state used to define the QCFs of this operator.  Because of its multiplicative renormalizability, matrix element of ${\cal O}_{q}^{\nu,b}$ in Eq.~(\ref{eq:qqo}) 
with any state could 
define an allowable renormalization scheme,
\begin{align}\label{eq:ZX}
\begin{split}
Z^{\text{RS}}(\xi^2, \mu^2,\delta)=
\frac{\langle \text{RS} | \hat{n} \cdot {\cal O}_{q}^b(\xi,\mu^2,\delta) | \text{RS} \rangle}
{ \langle \text{RS} | \hat{n} \cdot {\cal O}_{q}^b(\xi,\mu^2,\delta) | \text{RS} \rangle^{(0)}} \,,
\end{split}
\end{align}
where $\hat{n}$ is any vector keeping the denominator nonvanishing and the superscript ``(0)" indicates that the matrix element is evaluated to the leading order (LO) in perturbation theory.  Different choice of the state $|\text{RS}\rangle$ corresponds to different renormalization scheme. For example, an off-shell quark state with a specific momentum was used in defining RI/MOM or RI$^\prime$/MOM scheme \cite{Constantinou:2017sej,Alexandrou:2017huk,Chen:2017mzz,Stewart:2017tvs,Wang:2019tgg}; a hadron state with zero momentum was used in calculations of pseudo-PDFs~\cite{Orginos:2017kos} [Matrix element in this case cannot be perturbatively calculated and one
may choose the denominator in Eq.~\eqref{eq:ZX} as $1$]; and the vacuum state was introduced in Ref.~\cite{Braun:2018brg}.

In the following, we define the renormalization constant with the vacuum state $|\Omega\rangle$ and denote $\text{RS}=\text{vac}$. By calculating the vacuum expectation value to NNLO, we demonstrate that without an identified external momentum, the renormalization constant $Z^{\text{vac}}$ is completely free of infrared (IR) and collinear (CO) singularity and its UV divergence is regularized by DR, from which we obtain $Z^{\overline{\text{MS}}}(\xi^2,\mu^2,\epsilon)$ and $R^{\text{vac}}(\xi^2,\mu^2)$ at NNLO level.

\begin{figure}[thb]
	\begin{center}
		\includegraphics[width=3.3in]{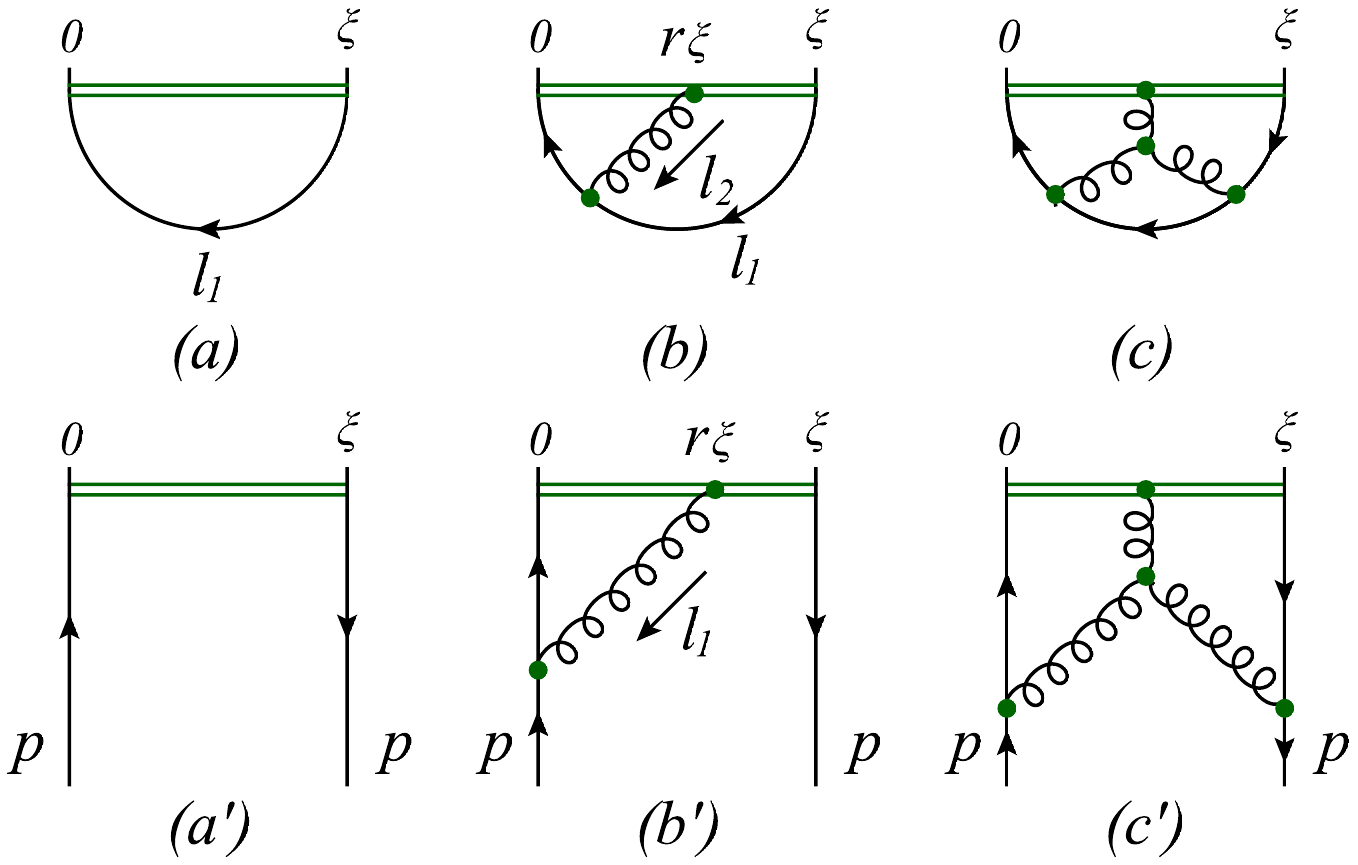}
		\caption{\label{fig:vacuum-quasi-quark-example}
		Representative Feynman diagrams, up to NNLO, for the vacuum expectation value of quark correlation operator (top row), and for the non-singlet quark matrix elements of the same operator (bottom row).}
	\end{center}
\end{figure}

In Fig.~\ref{fig:vacuum-quasi-quark-example}(a,b,c), we show some representative Feynman diagrams, up to NNLO, for the vacuum expectation $\langle \Omega| \hat{n} \cdot {\cal O}_{q}^b(\xi,\mu^2,\delta) | \Omega \rangle$.
The diagram (a) in Fig.~\ref{fig:vacuum-quasi-quark-example} determines the normalization of $Z^{\text{vac}}$,
\begin{align}\label{eq:Ma0}
\begin{split}
\langle \Omega | \hat{n} \cdot {\cal O}^{\text{b}}_{q} | \Omega \rangle^{(0)}=& \,2 N_c \,\mu^{4-d} \,\pi^{-d/2} \,\Gamma(d/2) \,|\xi|^{-d} \hat{n}\cdot \xi  \,,
\end{split}
\end{align}
where $|\xi|^2 \equiv -\xi^2$, and the result agrees with Ref. \cite{Braun:2018brg}.

The Fig.~\ref{fig:vacuum-quasi-quark-example}(b) is a representative Feynman diagram contributing to NLO $Z^{\text{vac}}$,
\begin{align}\label{eq:Mb1}
\begin{split}
M_b =& \,g_s^2 N_c C_F \,\mu^{8-2d} \int_0^{1}dr \int \frac{d^d l_1 \,d^d l_2}{(2\pi)^{2d}} \,e^{il_1 \cdot \xi +irl_2 \cdot \xi} \\
&\times \frac{\text{Tr}[ (\slashed{l}_1+\slashed{l}_2) \,\slashed{\xi} \,\slashed{l}_1 \,{\slashed{\hat{n}}} \,]}{(l_1^2+i0^+)(l_2^2+i0^+)((l_1+l_2)^2+i0^+)}\,,
\end{split}
\end{align}
where we assume without loss of generality that $z$-component $\xi_z$ is only nonzero component of $\xi$, and $\hat{n}$ satisfies $\hat{n} \cdot l \equiv l_z$ for any vector $l$. We find that it is convenient to carry out the integration in Eq.~(\ref{eq:Mb1}) by Fourier transforming the $\xi_z$ into $q_z$ in momentum space as $\mathscr{F} \left[ M_b \right] \equiv \int d\xi_z e^{-i \xi_z q_z} M_b$ to eliminate the exponential factor by using
\begin{align}\label{eq:FtMb11}
&\int d\xi_z e^{-i \xi_z q_z} \xi_z \int_0^{1}dr \,e^{-il_{1z} \xi_z -irl_{2z} \xi_z} \,,\\
=&2 i\, \text{Im}\left(\frac{1}{(q_z+l_{1z}+l_{2z}+i0^+)(q_z+l_{1z}+i0^+)}\right) ,
\nonumber
\end{align}
where $2\pi \delta(x)=-2\,\text{Im}(\frac{1}{x+i0^+})$ is used. The Fourier transformation also ensures that only imaginary part of gauge-link-related propagators are involved, which led to a similar effect of optical theorem. Our matrix element is defined with gauge-link in coordinate space, which is effectively equal to sum over diagrams with cut gauge-link in momentum space.  It is the summation of cuts of gauge link that forces the appearance of imaginary part of ``forward scattering amplitude". The obtained loop integrals in momentum space can be reduced to linear combination of a small set of integrals, called master integrals (MIs), by using integration-by-parts relations (IBPs)~\cite{Chetyrkin:1981qh,Laporta:2001dd}. We use the package \verb"FIRE5"~\cite{Smirnov:2014hma} to do this reduction, which results in
\begin{align}\label{eq:FtMb12}
\begin{split}
 \mathscr{F} \left[ M_b \right] &= i g_s^2 N_c C_F \,\mu^{8-2d} \, \frac{2(d-2)}{d-4} \\
& \times \left[ \ I_1 - \,\frac{2(2d-5)(3d-10)}{(d-3)(d-4)} \,q_z^{-1} I_2 \right]\,,
\end{split}
\end{align}
with two vacuum MIs defined as
\begin{align}\label{eq:MIv1}
\begin{split}
I_1 =& \int \frac{d^d l_1 \,d^d l_2}{(2\pi)^{2d}} \,\frac{1}{(l_1^2+i0^+)(l_2^2+i0^+)} \,\\
&\times 2\,\text{Im}\left(\frac{1}{(q_z+l_{1z}+i0^+)(q_z+l_{2z}+i0^+)}\right) \,, \\
I_2 =& \int \frac{d^d l_1 \,d^d l_2}{(2\pi)^{2d}} \,\frac{1}{(l_1^2+i0^+)(l_2^2+i0^+)}\\
&\times 2\,\text{Im}\left(\frac{1}{q_z+l_{1z}+l_{2z}+i0^+}\right) \,.
\end{split}
\end{align}
To carry out these single-scale vacuum MIs, we use the method presented in Ref.~\cite{Lee:2010wea} by setting up and solving dimensional recurrence relations and obtain
\begin{align}\label{eq:MIv1r}
\begin{split}
I_1 =&  \,\frac{\pi^{-d}}{8} \sin(d \pi) \Gamma(d/2-1)^2 \Gamma(3-d)^2 \,|q_z|^{2d-9} \,q_z^3 \,, \\
I_2 =&  \,\frac{\pi^{-d}}{8} \sin(d \pi) \Gamma(d/2-1)^2 \Gamma(5-2d) \,|q_z|^{2d-9} \,q_z^4 \,.
\end{split}
\end{align}
We then Fourier transform inversely from $q_z$ dependence into $\xi_z$ dependence to derive the result of $M_b$ in DR. Other two-loop diagrams, including UV counter term diagrams, can be calculated similarly.

All three-loop diagrams like diagram (c) in Fig.~\ref{fig:vacuum-quasi-quark-example} can also be calculated similarly as the diagram (b) described above. The only difference is that analytical expression of vacuum MIs cannot be obtained by solving dimensional recurrence relations directly. Instead, we calculate the vacuum MIs to high accuracy by using dimensional recurrence relations and then obtain exact results by using PSLQ algorithm \cite{Bailey:1999nv}. We check the correctness of our exact results numerically with at least $10^3$ digits.

By adding all diagrams and UV counter terms together, the remained divergences should be removed by operator renormalization. With a $\overline{\text{MS}}$ subtraction scheme, we obtain $Z^{\overline{\text{MS}}}(\xi^2,\mu^2,\epsilon)$ and $R^{\text{vac}}(\xi^2,\mu^2)$ at NNLO level, with analytical expressions given in supplementary material.

\sect{Matching coefficients}
\label{sec:hme}
By choosing the $\overline{\text{MS}}$ scheme for QCFs, we have the same
factorization in Eq.~(\ref{eq:factorization}) with $R^{\text{RS}}=1$,
{ which leads to}
a $\mu$ dependence on the left hand of the equation.  To calculate the matching coefficients $K^{\nu}$,
we replace the hadron $h$ in Eq.~(\ref{eq:factorization}) by a quark state and expand both sides perturbatively,
\begin{align}
F_{q_{ik}/q_i}^{\nu(n)}(\omega,\xi^2, \mu^2)
&= \sum_{m=0}^{n} \int_{-1}^1 \frac{dx}{x}\, f^{(m)}_{q_{ik}/q_i}(x,\mu^2)
\nonumber\\
&{\hskip 0.3in} \times
K^{\nu(n-m)}(x\omega,\xi^2, \mu^2)\,,
\label{eq:coef}
\end{align}
with $n,m=0,1,2$ indicating the power in $\alpha_s$.  While partonic $f^{(n)}_{q_{ik}}$ with $n=0,1,2$ in the $\overline{\rm MS}$ factorization scheme are known~\cite{Curci:1980uw}, we have to calculate partonic version of 
$F_{q_{ik}/q_i}^{\nu(n)}$ in the $\overline{\rm MS}$ scheme perturbatively to $n=0,1,2$ to derive the NNLO matching coefficient $K^{\nu(n)}$.

Some representative Feynman diagrams for $F_{q_{ik}/q_i}^{\nu(n)}$ are shown in Fig.~\ref{fig:vacuum-quasi-quark-example} (a$^\prime$, b$^\prime$, c$^\prime$). The diagram \ref{fig:vacuum-quasi-quark-example}(a$^\prime$) gives the tree level result
\begin{align}\label{eq:qqPDF0}
F_{q_{ik}/q_i}^{\nu(0)}= -2 \,i \,p^\nu e^{i\omega} \,.
\end{align}

To calculate $F_{q_{ik}/q_i}^{\nu}(\omega, \xi^2, \mu^2)$ at high orders, we again use transformation as Eq.~(\ref{eq:FtMb11})
to remove the exponential by going to momentum space, and then reduce the loop integrals to MIs by using IBPs. For example, at NLO we have two MIs:
\begin{align}\label{eq:MI1}
I_1^{(1)} =& \int \frac{d^d l_1}{(2\pi)^{d}} \,\frac{1}{l_1^2+i0^+}\,2\,\text{Im}\left(\frac{1}{q_z+l_{1z}+i0^+}\right) , \\
I_2^{(1)} =& \int \frac{d^d l_1}{(2\pi)^{d}} \,\frac{1}{l_1^2+i0^+}\,2\,\text{Im}\left(\frac{1}{q_z+l_{1z}+p_z+i0^+}\right)
\nonumber
\end{align}
and at NNLO we have 21 MIs.
The MIs generated from $n$-loop diagrams for $F_{q_{v}/q}^{\nu(n)}$ are functions satisfied
\begin{align}\label{eq:MITay}
I_j^{(n)}(y,p_z;d) = |q_z|^{d_n} q_z^{d_{nj}} \,J_j^{(n)}(y;d) \,,
\end{align}
where $y \equiv p_z/q_z$, $d_n \equiv -2n\epsilon-1$, and $d_n+d_{nj}$ are the dimensions of MI $I_j^{(n)}$. These MIs can be derived by solving the differential equations~\cite{Kotikov:1990kg}
\begin{align}\label{eq:DialE}
\partial_y J_j^{(n)}(y;d) = \sum_{k} A_{jk}(y;d) \,J_k^{(n)}(y;d) \,,
\end{align}
with $J_j^{(n)}(0;d)$ serving as boundary conditions. By applying IBPs again, the integrals in boundary conditions can be decomposed into vacuum MIs at $n$-loop order, which have been calculated in the renormalization procedure. Therefore, $J_j^{(n)}$ can be expanded as a Taylor series of $y$ based on the differential equations in Eq.~(\ref{eq:DialE}).

After carrying out MIs, we can Fourier transform back to position space and the $y$ dependence is changed to dependence on $\omega$. Analytical results can be obtained by fitting the Taylor series of $\omega$ with proper ansatz \cite{Moch:2004pa} in terms of harmonic polylogarithms~\cite{Goncharov:1998kja,Remiddi:1999ew,Borwein:1999js}. By adding contributions from all diagrams and then multiplying it by UV renormalization factor $Z^{-1}_{\overline{\text{MS}}}$, we obtain perturbative results of $F_{q_{ik}/q_i}^{\nu(n)}(\omega, \xi^2, \mu^2)$ with $n=1,2$.  We then obtain $\overline{\text{MS}}$ matching coefficients $K^{\nu(n)}(x\omega,  \xi^2, \mu^2)$ using
Eq.~(\ref{eq:coef}). As expected, all divergences are canceled and final results of $K^{\nu(n)}$ are finite. It verifies the proof of the factorization theorem~\cite{Ma:2014jla} up to two-loop order. Our one-loop results $K^{\nu(1)}$ agree with previous calculations~\cite{Ji:2017rah,Radyushkin:2017lvu,Izubuchi:2018srq}; terms proportional to $n_f$ in two-loop results have been calculated in Ref.~\cite{Braun:2018brg} using quark mass regulator; while other two-loop results are new.  By performing Fourier transformation, we also obtain analytical matching coefficients for pseudo-PDFs and quasi-PDFs. All analytical results are given in the supplementary material.

Using Eq.~(\ref{eq:factorization}), one can obtain NNLO matching coefficients in other $\text{RS}$ by calculating corresponding $R^{\text{RS}}$.

\begin{figure}[t]
	\begin{center}
		\includegraphics[width=3.3in]{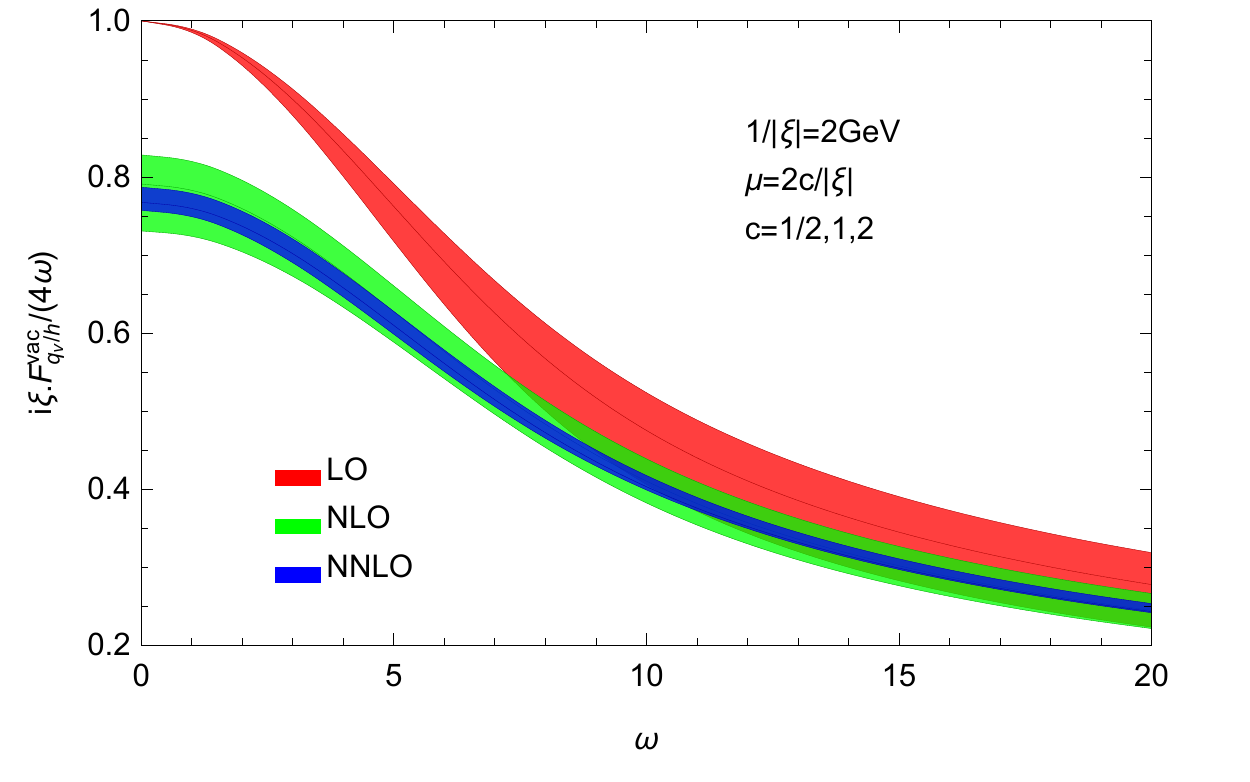}
		\includegraphics[width=3.3in]{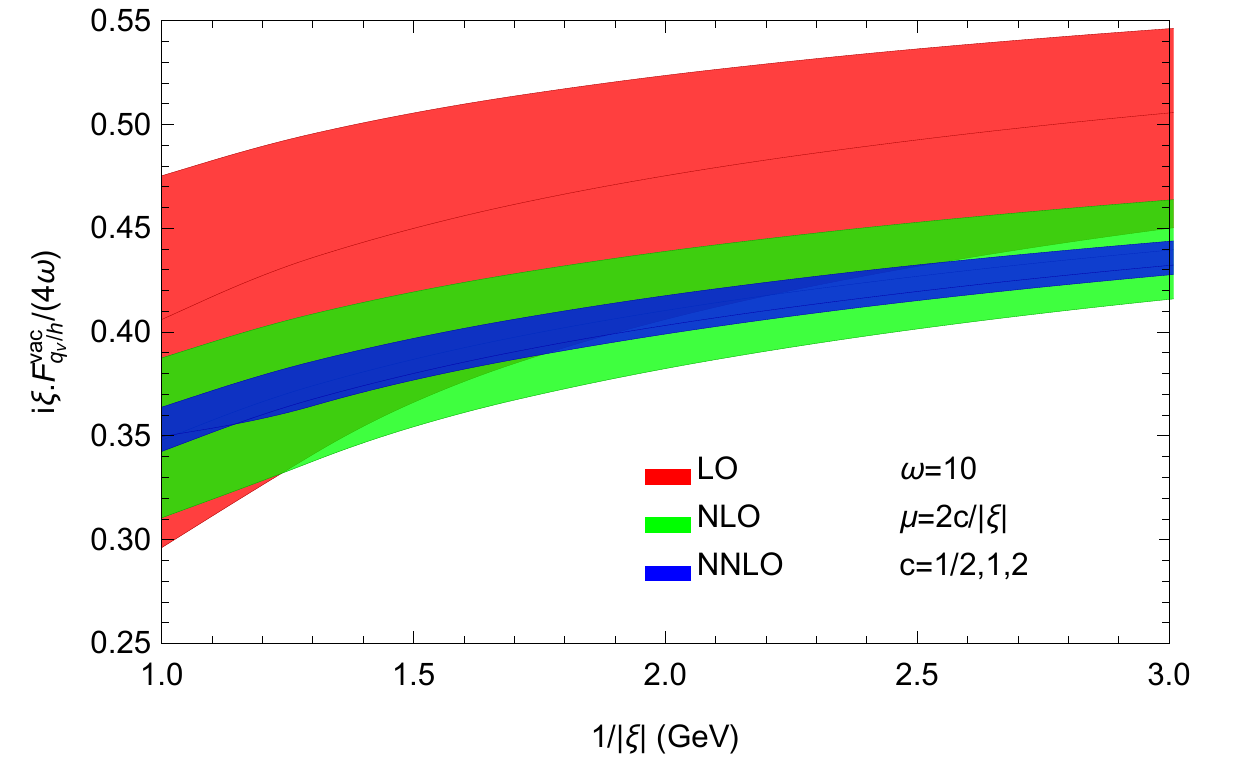}
		\caption{\label{fig:plotO}
		Numerical predictions for valence-quark correlation functions 
		with LO, NLO and NNLO matching coefficients and CT18NNLO PDFs. }
	\end{center}
\end{figure}

\sect{Numerical results}
\label{sec:num}
{ 
With our calculated matching coefficients at LO, NLO and NNLO and the factorization formula in Eq.~(\ref{eq:factorization}), we can predict valence-quark correlation functions 
by using existing PDFs extracted from experimental data, and test them by LQCD calculations.}
In Fig.~\ref{fig:plotO}, we present $\frac{i}{4\omega} \xi\cdot F_{q_{v}/h}^{\text{vac}}(\omega,\xi^2)$ as a function of $\omega$ with fixed $1/|\xi|=2$ GeV or as a function of $1/|\xi|$ with fixed $\omega=10$.
{
We used CT18NNLO PDFs \cite{Hou:2019efy}, and}
set $\mu=2c/|\xi|$ to minimize logarithms encountered in perturbative calculation. 
{ 
We chose $c=1$ for the predicted curves, and varied $c=1/2$ to $2$ for the bands}
to estimate theoretical uncertainties due to ambiguity of scale choice. 
{
Our numerical predictions show a great improvement in perturbative uncertainty when NNLO matching is used, 
especially, for the region where $1/|\xi|$ is small and more lattice data are available~\cite{Bringewatt:2020ixn}.
The}
NNLO results can reduce theoretical uncertainty by more than a factor of $3$ comparing with NLO results.

\sect{Summary}
\label{sec:summary}
Properly renormalized quark correlation functions in position space, if $\xi^2\Lambda_{\rm QCD}^2$ is sufficiently small,  are good LQCD observables that are calculable in LQCD and factorizable to PDFs~\cite{Ma:2014jla,Ma:2017pxb,Izubuchi:2018srq}.  We discussed the ambiguity and scheme-dependence 
{ of}
the multiplicative renormalization constant $Z^{\text{RS}}$, and demonstrated that $Z^{\text{RS}}$ defined with the vacuum state is advantageous for carrying out perturbative calculations of the matching coefficients, especially, at high order in $\alpha_s$.  For the first time, we derived a complete NNLO flavor non-singlet coefficient functions for QCFs, 
{ 
and predicted valence-quark correction functions in Fig.~\ref{fig:plotO} by using existing PDFs and our 
matching coefficients.  We clearly demonstrated that importance of NNLO matching coefficients for reducing the perturbative uncertainty in our factorization approach. 
Comparing our predictions with LQCD data will provide the first test of compatibility between LQCD calculations and high energy experimental measurements in terms of QCD factorization at the NNLO accuracy~\cite{Bringewatt:2020ixn}.

Our definition of QCFs and method of calculations can be easily generalized to gluon correlation functions (GCFs). 
With multiple ``good" LQCD observables, including the QCFs and GCFs, as well as the current-current correlation functions (better UV behavior) \cite{Ma:2017pxb}, and our ability to calculate NNLO matching coefficients, the extraction of PDFs from LQCD calculations in terms of QCD factorization approach can be in fact at the same rigor as how PDFs have been extracted from experimental data.  In addition to the complementary revenue for extracting PDFs or other partonic structures of hadrons, LQCD calculation provides}
a tremendous potential to extract partonic structure of hadrons that could be difficult to do scattering experiments with.

\section*{Acknowledgments}
\label{sec:acknowledgments}

We thank L. Leskovec, R.Sufian and Y.-B. Yang for useful discussions.
This work of Z.-Y.L. and Y.-Q.M. is supported by the National Natural Science Foundation of China (Grants No. 11875071, No. 11975029) and the High-performance Computing Platform of Peking University, and J.-W.Q. is supported by the U.S. Department of Energy contract DE-AC05-06OR23177, under which Jefferson Science Associates, LLC, manages and operates Jefferson Lab.


{\textbf{Note added:} } Recently, some related preprints appeared \cite{Braun:2020ymy,Chen:2020arf,Chen:2020iqi}. In Ref.~\cite{Braun:2020ymy} the authors obtained NNLO results for $Z^{\overline{\text{MS}}}$ and $R^{\text{vac}}$, which exactly agree with our results. In Refs.~\cite{Chen:2020arf,Chen:2020iqi} the authors obtained matching coefficients for flavor non-diagonal quark to quark channel that starts from two-loop order. Reference \cite{Chen:2020ody}, where flavor non-singlet matching coefficients for quasi-PDF are also calculated to NNLO, includes results that are in agreement with ours.

\appendix
\begin{widetext}
	
	\section*{A: Perturbative results for quark correlation function}\label{sec:QCF}
	
	Renormalization factor in $\overline{\text{MS}}$ subtraction scheme is obtained as
	\begin{align}
	& Z_{\overline{\text{MS}}} =1+ \frac{\alpha_s S_\epsilon}{\pi {\epsilon}}  C_F +\left(\frac{\alpha_s S_\epsilon}{\pi {\epsilon}}\right)^2 \!\!   C_F \Bigg\{ \bigg[ \frac{C_F}{2} -\frac{13 C_A}{32} +\frac{n_f T_F}{8}  \bigg] + \bigg[\Big( \,\frac{\pi^2}{12} -\frac{1}{8} \Big) \,C_F - \Big( \frac{\pi^2}{48} -\frac{25}{48} \Big) \,C_A  -\frac{n_f T_F}{6} \bigg] {\epsilon} \Bigg\} \,,
	\label{eq:Zmq}
	\end{align}
	where $S_\epsilon \equiv (4\pi)^{\epsilon} /\Gamma (1-\epsilon) $ is a conventional factor in the $\overline{\text{MS}}$ scheme.
	
	The finite renormalization factor $R^{\text{vac}}(\xi^2,\mu^2)$ is obtained as
	\begin{align}\label{eq:ZmqZv}
	\begin{split}
	& R^{\text{vac}}(\xi^2,\mu^2)=1 +\frac{\alpha_s }{\pi} C_F \Bigg( \frac{3}{4} L +\frac{\pi^2}{3} +2 \Bigg) +\frac{\alpha_s^2}{\pi^2} C_F \Bigg\{ \bigg[\frac{9}{32} C_F+ \frac{11}{32} C_A  -\frac{1}{8} n_f T_F \bigg] L^2 \\
	&\, +\bigg[ \Big( \,\frac{5\pi^2}{12}+\frac{43}{32} \Big) \,C_F + \Big( \,\frac{19\pi^2}{72}+\frac{75}{32} \Big) \,C_A  -\Big( \,\frac{\pi^2}{9}+\frac{7}{8} \Big) \,n_f T_F \bigg] L +\bigg[ \Big( \,\frac{\pi^4}{90}-\frac{\zeta_3}{2}+\frac{13\pi^2}{12}+\frac{153}{128} \Big) \,C_F \\
	&\, - \Big( \,\frac{\pi^4}{90}+\frac{13\zeta_3}{2}+\frac{5\pi^2}{432}-\frac{6413}{1152} \Big) \,C_A + \Big( \,2\zeta_3+\frac{\pi^2}{27}-\frac{589}{288} \Big) \,n_f T_F \bigg] \Bigg\} \,,
	\end{split}
	\end{align}
	where $L\equiv \ln(-\xi^2 \mu^2/4)+2\gamma_E$.
	
	We express
	$K^\nu(x\omega, \xi^2, \mu^2) \equiv x p^\nu \,A(x\omega, \xi^2, \mu^2)   + \,x \omega \frac{\xi^\nu}{-\xi^2} \,B(x\omega, \xi^2, \mu^2) $, and as argued in Ref.~\cite{Ma:2017pxb}, $A(\omega, \xi^2, \mu^2)$ and $B(\omega, \xi^2, \mu^2)$ are analytical functions of $\omega$ everywhere except infinity.
	$A(\omega, \xi^2, \mu^2)$ and $B(\omega, \xi^2, \mu^2)$ can be decomposed as
	\begin{align}\label{eq:ABdec}
	\begin{split}
	i A(\omega, \xi^2, \mu^2) &= 2 \,e^{i\omega} +\frac{\alpha_s }{\pi} \left\{ \sum_{i=0}^{1} L^i C_1^{(1)} \,a_{i10}^{(1)} \,e^{i\omega} +\sum_{i=0}^{1} L^i C_1^{(1)} \int_0^1 dz  a_{i11}^{(1)} (z) \left( e^{iz\omega} -e^{i\omega} \right)  \right\}  \\
	& {\hskip -0.8in} + \frac{\alpha_s^2}{\pi^2} \left\{ \sum_{i=0}^{2} \sum_{j=1}^{3} L^i C_j^{(2)} a_{ij0}^{(2)} \,e^{i\omega}  +\sum_{i=0}^{2} \sum_{j=1}^{3} L^i C_j^{(2)} \int_0^1 dz \,a_{ij1}^{(2)} (z) \left( e^{iz\omega} -e^{i\omega} \right)  +\sum_{i=0}^{1} L^i C_4^{(2)} \int_{-1}^0 dz \,a_{i42}^{(2)} (z) \left( e^{iz\omega} -e^{i\omega} \right)  \right\} \,, \end{split}  \\
	\begin{split}
	i B(\omega, \xi^2, \mu^2) &= \frac{\alpha_s }{\pi} \left\{ C_1^{(1)} b_{010}^{(1)} \,e^{i\omega} +C_1^{(1)} \int_0^1 dz \,b_{011}^{(1)} (z) \left( e^{iz\omega} -e^{i\omega} \right)  \right\}  \\
	& {\hskip -0.8in} + \frac{\alpha_s^2}{\pi^2} \left\{ \sum_{i=0}^{1} \sum_{j=1}^{3} L^i C_j^{(2)} b_{ij0}^{(2)} \,e^{i\omega}  +\sum_{i=0}^{1} \sum_{j=1}^{3} L^i C_j^{(2)} \int_0^1 dz \,b_{ij1}^{(2)} (z) \left( e^{iz\omega} -e^{i\omega} \right)  +C_4^{(2)} \int_{-1}^0 dz \,b_{042}^{(2)} (z) \left( e^{iz\omega} -e^{i\omega} \right)  \right\}  \,,
	\end{split}
	\end{align}
	where $C_1^{(1)} \equiv C_F$, $C_1^{(2)} \equiv C_F^2$, $C_2^{(2)} \equiv C_F C_A$, $C_3^{(2)} \equiv n_f C_F T_F$, $C_4^{(2)} \equiv C_F^2 -\frac{1}{2} C_F C_A$.
	Analytical expressions of $a_{ijk}^{(n)}$ and $b_{ijk}^{(n)}$ are give by
	\begin{align}\label{eq:A1B1A2B2}
	&a_{110}^{(1)} = \,\frac{3}{2} \,,
	{\hskip 0.3in} a_{111}^{(1)}(z) = \frac{z^2+1}{z-1} \,,
	{\hskip 0.3in} a_{010}^{(1)} = \,\frac{5}{2} \,,
	{\hskip 0.3in} a_{011}^{(1)}(z) = -\frac{z^2-4z+1}{z-1} -\frac{4\text{H}(1;z)}{z-1} \,, \nonumber \\
	& a_{210}^{(2)} = \,\frac{9}{16} \,,
	{\hskip 0.3in} a_{211}^{(2)}(z) = \,\frac{z-1}{2} +\frac{(3z^2+1)\text{H}(0;z)}{4(z-1)} +\frac{(z^2+1)\text{H}(1;z)}{z-1}  \nonumber  \,, \\
	& a_{220}^{(2)} = \,\frac{11}{16} \,,
	{\hskip 0.3in} a_{221}^{(2)}(z) = \,\frac{11(z^2+1)}{24(z-1)} \,,
	{\hskip 0.3in} a_{230}^{(2)} = -\frac{1}{4} \,,
	{\hskip 0.3in} a_{231}^{(2)}(z) = -\frac{z^2+1}{6(z-1)}  \nonumber  \,, \\
	& a_{110}^{(2)} = \,\frac{\pi^2}{3} +\frac{25}{16} \,,
	{\hskip 0.3in} a_{111}^{(2)}(z) = \,\frac{(z^2+1)\pi^2}{3(z-1)} -\frac{7(z-1)}{2} -\frac{(z^2-6z+4)\text{H}(0;z)}{2(z-1)} -\frac{2(z^2-4z+1)\text{H}(1;z)}{z-1} \nonumber \\
	& \ +\frac{(z+1)\text{H}(0,0;z)}{2} -\frac{(z^2+1)\text{H}(0,1;z)}{z-1} +\frac{(z^2-3)\text{H}(1,0;z)}{z-1} -\frac{2(z^2+5)\text{H}(1,1;z)}{z-1}  \nonumber  \,, \\
	& a_{120}^{(2)} = -\frac{\pi^2}{12} +\frac{53}{16} \,,
	{\hskip 0.3in} a_{121}^{(2)}(z) = -\frac{(z^2+1)\pi^2}{12(z-1)} +\frac{77z^2-54z+77}{18(z-1)} +\frac{(5z^2+17)\text{H}(0;z)}{12(z-1)} -\frac{11\text{H}(1;z)}{3(z-1)} \nonumber \\
	& \ +\frac{(z^2+1)\text{H}(0,0;z)}{2(z-1)} \,,
	{\hskip 0.3in} a_{130}^{(2)} = -\frac{5}{4} \,,
	{\hskip 0.3in} a_{131}^{(2)}(z) = -\frac{8(z^2+1)}{9(z-1)} -\frac{(z^2+1)\text{H}(0;z)}{3(z-1)} +\frac{4\text{H}(1;z)}{3(z-1)} \,, \nonumber \\
	& a_{142}^{(2)}(z) = \,\frac{(z^2+1)\pi^2}{6(z-1)} +2(z+1) -(z-1)\text{H}(0;-z) +\frac{2(z^2+1)\text{H}(-1,0;-z)}{z-1} -\frac{(z^2+1)\text{H}(0,0;-z)}{z-1} \,, \nonumber \\
	& a_{010}^{(2)} = -4\zeta_3 +\frac{\pi^2}{9} +\frac{223}{192} \,,
	{\hskip 0.3in} a_{011}^{(2)}(z) = -\frac{2(5z^2+4)\zeta_3}{z-1} -\frac{(3z^3+z^2-8z+3)\pi^2}{6(z-1)} +\frac{(z^2-7)\pi^2\text{H}(1;z)}{6(z-1)}  \nonumber \\
	& \ +\frac{17z^2-30z+17}{4(z-1)} -\frac{(24z^2-23z+5)\text{H}(0;z)}{4(z-1)} +\frac{(z^2+11z-4)\text{H}(1;z)}{4(z-1)} +2(z+1)\text{H}(-1,0;z)  \nonumber \\
	& \ -\frac{(4z^3+10z^2-10z+3)\text{H}(0,0;z)}{4(z-1)} -\frac{(2z^3-5z^2+10z-1)\text{H}(0,1;z)}{2(z-1)} -\frac{(2z^3+5z^2-12z+3)\text{H}(1,0;z)}{2(z-1)}  \nonumber \\
	& \ -\frac{(z^3-3z^2+9z+5)\text{H}(1,1;z)}{z-1} -\frac{2(z^2+1)\text{H}(0,-1,0;z)}{z-1} +\frac{(z+1)\text{H}(0,0,0;z)}{4} -\frac{2(z^2+1)\text{H}(0,1,0;z)}{z-1}  \nonumber \\
	& \ +\frac{(z^2+1)\text{H}(0,1,1;z)}{2(z-1)} -\frac{(z^2+1)\text{H}(1,0,0;z)}{2(z-1)} -2(z+1)\text{H}(1,0,1;z) -\frac{(z^2-3)\text{H}(1,1,0;z)}{z-1} +\frac{24\text{H}(1,1,1;z)}{z-1} \,,  \nonumber \\
	& a_{020}^{(2)} = \,\zeta_3 -\frac{5\pi^2}{24} +\frac{4877}{576} \,,
	{\hskip 0.3in} a_{021}^{(2)}(z) = \,\frac{(7z^2+3)\zeta_3}{4(z-1)} -\frac{(2z^2-1)\pi^2}{24(z-1)} -\frac{z^2\pi^2\text{H}(1;z)}{3(z-1)} -\frac{77z^2-1104z+77}{108(z-1)}  \nonumber \\
	& \ +\frac{(101z^2+3z+29)\text{H}(0;z)}{36(z-1)} -\frac{(27z^2-36z+409)\text{H}(1;z)}{36(z-1)} -(z+1)\text{H}(-1,0;z) +\frac{(17z^2+24z+11)\text{H}(0,0;z)}{24(z-1)}  \nonumber \\
	& \ -\frac{(z^2-2z)\text{H}(0,1;z)}{4(z-1)} +\frac{(3z^2+6z-53)\text{H}(1,0;z)}{12(z-1)} -\frac{(3z^2-6z-85)\text{H}(1,1;z)}{12(z-1)} +\frac{(z^2+1)\text{H}(0,-1,0;z)}{z-1}  \nonumber \\
	& \ +\frac{(z^2+1)\text{H}(0,0,0;z)}{4(z-1)} -\frac{(z^2+1)\text{H}(0,1,0;z)}{2(z-1)} -\frac{2\text{H}(1,0,0;z)}{z-1} +\frac{(z^2+1)\text{H}(1,0,1;z)}{2(z-1)} -\frac{3(z^2+1)\text{H}(1,1,0;z)}{2(z-1)} \,,  \nonumber \\
	& a_{030}^{(2)} = -\frac{469}{144} \,,
	{\hskip 0.3in} a_{031}^{(2)}(z) = -\frac{19(z^2+6z+1)}{54(z-1)} -\frac{(5z^2+12z+5)\text{H}(0;z)}{18(z-1)} +\frac{32\text{H}(1;z)}{9(z-1)} -\frac{(z^2+1)\text{H}(0,0;z)}{6(z-1)}  \nonumber \\
	& \ +\frac{4\text{H}(1,0;z)}{3(z-1)} -\frac{8\text{H}(1,1;z)}{3(z-1)}  \,,  \nonumber \\
	& a_{042}^{(2)}(z) = -\frac{(7z^2+3)\zeta_3}{2(z-1)} -\frac{(z^2-6z+1)\pi^2}{12(z-1)} +\frac{(z^2+5)\pi^2\text{H}(-1;-z)}{6(z-1)} -\frac{21(z+1)}{4} +\frac{(19z^2-18z+3)\text{H}(0;-z)}{4(z-1)}  \nonumber \\
	& \ +\frac{(z^2+6z-3)\text{H}(-1,0;-z)}{z-1} -\frac{2(2z^2-z)\text{H}(0,0;-z)}{z-1} -(z+1)\text{H}(1,0;-z) +\frac{2(z^2+5)\text{H}(-1,-1,0;-z)}{z-1}  \nonumber \\
	& \ +\frac{(3z^2-1)\text{H}(-1,0,0;-z)}{z-1} -\frac{2(z^2+1)\text{H}(0,-1,0;-z)}{z-1} -\frac{(z^2+1)\text{H}(0,0,0;-z)}{2(z-1)} +\frac{(z^2+1)\text{H}(0,1,0;-z)}{z-1}  \,,  \nonumber\\
	& b_{010}^{(1)} = -1 \,,
	{\hskip 0.3in} b_{011}^{(1)}(z) = 2(z-1) \,,  \nonumber\\
	& b_{110}^{(2)} = -\frac{3}{4} \,,
	{\hskip 0.3in} b_{111}^{(2)}(z) = \,2(z-1) +(z-1)\text{H}(0;z) +2(z-1)\text{H}(1;z)  \nonumber  \,, \\
	& b_{120}^{(2)} = -\frac{11}{12} \,,
	{\hskip 0.3in} b_{121}^{(2)}(z) = \,\frac{11(z-1)}{6} \,,
	{\hskip 0.3in} b_{130}^{(2)} = \,\frac{1}{3} \,,
	{\hskip 0.3in} b_{131}^{(2)}(z) = -\frac{2(z-1)}{3}  \nonumber  \,, \\
	& b_{010}^{(2)} = -\frac{2\pi^2}{9} +\frac{17}{24} \,,
	{\hskip 0.3in} b_{011}^{(2)}(z) = -\frac{(3z^2-4z+2)\pi^2}{6} -12(z-1) +(3z+7)\text{H}(0;z) -(z-1)\text{H}(1;z) \nonumber \\
	& \ -(z^2+3z-1)\text{H}(0,0;z) -z^2\text{H}(0,1;z) -(z^2+2z-1)\text{H}(1,0;z) -(z^2+2z-3)\text{H}(1,1;z)  \nonumber  \,, \\
	& b_{020}^{(2)} = \,\frac{\pi^2}{6} -\frac{89}{72} \,,
	{\hskip 0.3in} b_{021}^{(2)}(z) = \,\frac{\pi^2}{6} +\frac{139(z-1)}{18} -\frac{(z+23)\text{H}(0;z)}{6} -(z-1)\text{H}(1;z) +(2z-1)\text{H}(0,0;z) \nonumber \\
	& \ +z\text{H}(1,0;z) \,,
	{\hskip 0.3in} b_{030}^{(2)} = \,\frac{1}{18} \,,
	{\hskip 0.3in} b_{031}^{(2)}(z) = -\frac{10(z-1)}{9} -\frac{2(z-1)\text{H}(0;z)}{3} \,, \nonumber \\
	& b_{042}^{(2)}(z) = \,\frac{(2z-1)\pi^2}{3} +6(z+1) -2(z-2)\text{H}(0;-z) +4(2z-1)\text{H}(-1,0;-z) -2(3z-1)\text{H}(0,0;-z) \,,
	\end{align}
	where the definition of harmonic polylogarithms $\text{H}$ can be found in Refs.~\cite{Goncharov:1998kja,Remiddi:1999ew,Borwein:1999js}. All harmonic polylogarithms encountered in this work can be expressed explicitly as {(with $0 < z < 1$)}
	\begin{align}\label{eq:HPL}
	& \text{H}(-1;z) = \,\ln(1+z) \,,
	{\hskip 0.3in} \text{H}(0;z) = \,\ln(z) \,,
	{\hskip 0.3in} \text{H}(1;z) = -\ln(1-z) \,,
	{\hskip 0.3in} \text{H}(-1,0;z) = \,\ln(z)\ln(1+z) +\text{Li}_2\left(-z\right) \,, \nonumber  \\
	& \text{H}(0,0;z) = \,\frac{1}{2}\ln^2(z) \,,
	{\hskip 0.3in} \text{H}(0,1;z) = \,\text{Li}_2\left(z\right) \,,
	{\hskip 0.3in} \text{H}(1,0;z) = -\ln(1-z)\ln(z) -\text{Li}_2\left(z\right) \,,  \nonumber \\
	& \text{H}(1,1;z) = \,\frac{1}{2}\ln^2(1-z) \,,
	{\hskip 0.3in} \text{H}(-1,-1,0;z) = \,\zeta_3 -\frac{\pi^2}{6}\ln(1+z) +\frac{1}{6}\ln^3(1+z) -\text{Li}_3\left(\frac{1}{1+z}\right) \,, \nonumber \\
	& \text{H}(-1,0,0;z) = \,\frac{1}{2}\ln^2(z)\ln(1+z) +\ln(z)\text{Li}_2\left(-z\right) -\text{Li}_3\left(-z\right) \,, \nonumber \\
	& \text{H}(0,-1,0;z) = -\ln(z)\text{Li}_2\left(-z\right) +2\text{Li}_3\left(-z\right) \,,
	{\hskip 0.3in} \text{H}(0,0,0;z) = \,\frac{1}{6}\ln^3(z) \,,
	{\hskip 0.3in} \text{H}(0,1,0;z) = \,\ln(z)\text{Li}_2\left(z\right) -2\text{Li}_3\left(z\right) \,, \nonumber \\
	& \text{H}(0,1,1;z) = \,\zeta_3 +\frac{1}{2}\ln^2(1-z)\ln(z) +\ln(1-z) \text{Li}_2\left(1-z\right) -\text{Li}_3\left(1-z\right) \,, \nonumber \\
	& \text{H}(1,0,0;z) = -\frac{1}{2}\ln(1-z)\ln^2(z) -\ln(z) \text{Li}_2\left(z\right) +\text{Li}_3\left(z\right) \,, \nonumber \\
	& \text{H}(1,0,1;z) = -2\zeta_3 -\frac{\pi^2}{3} \ln(1-z) +\ln^2(1-z)\ln(z) +\ln(1-z) \text{Li}_2\left(z\right) +2\text{Li}_3\left(1-z\right) \,, \nonumber \\
	& \text{H}(1,1,0;z) = \,\zeta_3 +\frac{\pi^2}{6} \ln(1-z) -\text{Li}_3\left(1-z\right) \,,
	{\hskip 0.3in} \text{H}(1,1,1;z) = -\frac{1}{6}\ln^3(1-z) \,. 
	\end{align}
	
	Complete analytical expressions of $a_{ijk}^{(n)}$ and $b_{ijk}^{(n)}$ and a code to calculate $i A(\omega, \xi^2, \mu^2)$ and $i B(\omega, \xi^2, \mu^2)$ are available to download from an ancillary file.
	
	\section*{B: Matching coefficients for pseudo-PDFs}\label{sec:ppdf}

	Pseudo-quark PDFs $\hat{F}_{q/h}^{\sigma,\text{RS}}$ can be obtained from quark correlation functions by performing Fourier transformations:
	\begin{align}\label{eq:pPDFs}
	& \hat{F}_{q/h}^{\sigma,\text{RS}} \left( y, \xi^2 \right)
	= \int \frac{d\omega}{2\pi} \frac{1}{p^\sigma} F_{q/h}^{\sigma,\text{RS}} \left(\omega, \xi^2 \right) e^{-iy\omega} \,,
	\end{align}
	where $\sigma$ can be  $t$ or $z$.
	The factorization formula in flavor non-singlet case is
	\begin{align}\label{eq:pPDFsfac}
	& \hat{F}_{q_{ik}/h}^{\sigma,\text{RS}} \left( y, \xi^2 \right)
	= \frac{1}{R^{\text{RS}}(\xi^2,\mu^2)} \int_{-1}^1 \frac{dx}{|x|} \, f_{q_{ik}/h}(x,\mu^2)
	\hat{K}^\sigma \left( y/x, \xi^2, \mu^2 \right)
	+{ O}(\xi^2\Lambda_{\rm QCD}^2) \,,
	\end{align}
	which results in
	\begin{align}\label{eq:MachpPDFs}
	& \hat{K}^{\sigma,\text{RS}} \left( y, \xi^2 \right)
	= \int \frac{d\omega}{2\pi} \frac{1}{p^\sigma} K^{\sigma,\text{RS}} \left(\omega, \xi^2 \right) e^{-iy\omega} \,.
	\end{align}

	The matching coefficients of pseudo-quark PDFs are obtained as
	\begin{align}\label{eq:pKdec}
	\begin{split}
	& i \hat{K}^t \left( y, \xi^2, \mu^2 \right) = 2\delta(y-1) +\frac{\alpha_s }{\pi} \left\{ \sum_{i=0}^{1} L^i C_1^{(1)} a_{i10}^{(1)} \delta(y-1) +\sum_{i=0}^{1} L^i C_1^{(1)} \int_0^1 dz \,a_{i11}^{(1)} (z) \left[ \delta(y-z) -\delta(y-1) \right] \right\} \\
	& {\hskip -0.1in} + \frac{\alpha_s^2}{\pi^2} \left\{ \sum_{i=0}^{2} \sum_{j=1}^{3} L^i C_j^{(2)} a_{ij0}^{(2)} \delta(y-1) +\left[ \sum_{i=0}^{2} \sum_{j=1}^{3} L^i C_j^{(2)} \int_0^1 dz  \,a_{ij1}^{(2)} (z)  +\sum_{i=0}^{1} L^i C_4^{(2)} \int_{-1}^0 dz a_{i42}^{(2)} (z) \right] \left[ \delta(y-z) -\delta(y-1) \right]   \right\} \,,  \end{split}  \\
	\begin{split}
	& i \hat{K}^t \left( y, \xi^2, \mu^2 \right) -i \hat{K}^z \left( y, \xi^2, \mu^2 \right) = \frac{\alpha_s }{\pi} \left\{ C_1^{(1)} b_{010}^{(1)} \delta(y-1) +C_1^{(1)} \int_0^1 dz \,b_{011}^{(1)} (z) \left[ \delta(y-z) -\delta(y-1) \right]  \right\}  \\
	& {\hskip -0.1in} + \frac{\alpha_s^2}{\pi^2} \left\{ \sum_{i=0}^{1} \sum_{j=1}^{3} L^i C_j^{(2)} b_{ij0}^{(2)} \delta(y-1) +\left[ \sum_{i=0}^{1} \sum_{j=1}^{3} L^i C_j^{(2)} \int_0^1 dz  \,b_{ij1}^{(2)} (z) +C_4^{(2)} \int_{-1}^0 dz b_{042}^{(2)} (z) \right] \left[ \delta(y-z) -\delta(y-1) \right]   \right\}  \,.
	\end{split}
	\end{align}

	\section*{C: Matching coefficients for quasi-PDFs}\label{sec:quasi}
	
	Quasi-quark PDFs $\tilde{F}_{q/h}^{\sigma,\text{RS}}$  can be obtained from quark correlation functions by performing Fourier transformation:
	\begin{align}\label{eq:qpPDFs}
	& \tilde{F}_{q/h}^{\sigma,\text{RS}} \left( y, 1/p_z^2 \right)
	= \int \frac{d\omega}{2\pi} \frac{1}{p^\sigma} F_{q/h}^{\sigma,\text{RS}} \left(\omega, -\omega^2/p_z^2 \right) e^{-iy\omega} \,,
	\end{align}
	where $\sigma$ can be  $t$ or $z$.
	The factorization formula in flavor non-singlet case is
	\begin{align}\label{eq:qpPDFsfac}
	& \tilde{F}_{q_{ik}/h}^{\sigma,\text{RS}} \left( y, 1/p_z^2 \right)
	= \frac{1}{R^{\text{RS}}(\xi^2,\mu^2)} \int_{-1}^1 \frac{dx}{|x|} \, f_{q_{ik}/h}(x,\mu^2)
	\tilde{K}^\sigma \left( y/x, 1/x^2p_z^2, \mu^2 \right)
	+{ O}\left( \Lambda_{\rm QCD}^2/(y p_z)^2 \right) \,,
	\end{align}
	which results in
	\begin{align}\label{eq:MachqpPDFs}
	& \tilde{K}^{\sigma,\text{RS}} \left( y, 1/p_z^2 \right)
	= \int \frac{d\omega}{2\pi} \frac{1}{p^\sigma} K^{\sigma,\text{RS}} \left(\omega, -\omega^2/p_z^2 \right) e^{-iy\omega} \,.
	\end{align}
	
	To have compact expressions, we define more plus functions as
	\begin{align}\label{eq:plusf}
	& [g(y)]_{\oplus (1)}^{[y_0]} \equiv \lim_{\lambda \to 1^+} \theta(y-\lambda) g(y) +\delta(y-\lambda) G(y,y_0) \,,
	{\hskip 0.3in} [g(y)]_{+ (1)}^{[y_0]} \equiv \lim_{\lambda \to 1^-} \theta(\lambda-y) g(y) -\delta(y-\lambda) G(y,y_0) \,, \nonumber \\
	& [g(y)]_{\oplus (1,\infty)}^{[y_0]} \equiv \lim_{\lambda \to 1^+ \atop \Lambda \to \infty} \theta(y-\lambda) \theta(\Lambda-y) g(y) + \left[ \delta(y-\lambda) -\delta(y-\Lambda) \right] G(y,y_0) \,, \\
	& [g(y)]_{+ (1,-\infty)}^{[y_0]} \equiv \lim_{\lambda \to 1^- \atop \Lambda \to -\infty} \theta(\lambda-y) \theta(y-\Lambda) g(y) - \left[ \delta(y-\lambda) -\delta(y-\Lambda) \right] G(y,y_0) \,, \nonumber
	\end{align}
	with $G(y,y_0) \equiv \int_{y_0}^y dy^\prime g(y^\prime)$.

	The matching coefficients of quasi-quark PDFs are obtained as
	\begin{align}\label{eq:qKdec}
	\begin{split}
	& i \tilde{K}^t \left( y, 1/p_z^2, \mu^2 \right) = 2\delta(y-1) +\frac{\alpha_s }{\pi} \left\{ \sum_{i=0}^{1} L_{\delta,i}(y) C_1^{(1)} a_{i10}^{(1)} +C_1^{(1)} \left[ \theta(y-1) \tilde{a}_{011}^{(1)} (y) \right]_{\oplus (1)}^{[\infty]} \right. \\
	& {\hskip -0.1in} \left. +\sum_{i=0}^{1} L_p^i C_1^{(1)} \left[ \theta(1-y) \theta(y) \tilde{a}_{i12}^{(1)} (y) \right]_{+ (1)}^{[-\infty]} +C_1^{(1)} \left[ \theta(-y) \theta(y+1) \tilde{a}_{013}^{(1)} (y) \right]_{+ (1)}^{[-\infty]} +C_1^{(1)} \left[ \theta(-y-1) \tilde{a}_{014}^{(1)} (y) \right]_{+ (1)}^{[-\infty]}  \right\}  \\
	& {\hskip -0.1in} + \frac{\alpha_s^2}{\pi^2} \left\{ \sum_{i=0}^{2} \sum_{j=1}^{3} L_{\delta,i}(y) C_j^{(2)} a_{ij0}^{(2)}  +\sum_{i=0}^{1} \sum_{j=1}^{3} L_p^i C_j^{(2)} \left[ \theta(y-1) \tilde{a}_{ij1}^{(2)} (y) \right]_{\oplus (1)}^{[\infty]} +\sum_{i=0}^{2} \sum_{j=1}^{3} L_p^i C_j^{(2)}  \left[ \theta(1-y) \theta(y) \tilde{a}_{ij2}^{(2)} (y) \right]_{+ (1)}^{[-\infty]}  \right. \\
	& {\hskip -0.1in} \left. +\sum_{i=0}^{1} \sum_{j=1}^{3} L_p^i C_j^{(2)} \left[ \theta(-y) \theta(y+1) \tilde{a}_{ij3}^{(2)} (y) \right]_{+ (1)}^{[-\infty]} +\sum_{i=0}^{1} \sum_{j=1}^{3} L_p^i C_j^{(2)} \left[ \theta(-y-1) \tilde{a}_{ij4}^{(2)} (y) \right]_{+ (1)}^{[-\infty]}  \right\} \,, \end{split}  \\
	\begin{split}
	& i \tilde{K}^t \left( y, 1/p_z^2, \mu^2 \right) -i \tilde{K}^z \left( y, 1/p_z^2, \mu^2 \right) = \frac{\alpha_s }{\pi} \left\{ L_{\delta,0}(y) C_1^{(1)} b_{010}^{(1)} +C_1^{(1)} \left[ \theta(1-y) \theta(y) \tilde{b}_{012}^{(1)} (y) \right]_{+ (1)}^{[-\infty]}  \right\}  \\
	& {\hskip -0.1in} + \frac{\alpha_s^2}{\pi^2} \left\{ \sum_{i=0}^{1} \sum_{j=1}^{3} L_{\delta,i}(y) C_j^{(2)} b_{ij0}^{(2)}  +\sum_{j=1}^{3} C_j^{(2)} \left[ \theta(y-1) \tilde{b}_{0j1}^{(2)} (y) \right]_{\oplus (1)}^{[\infty]} +\sum_{i=0}^{1} \sum_{j=1}^{3} L_p^i C_j^{(2)}  \left[ \theta(1-y) \theta(y) \tilde{b}_{ij2}^{(2)} (y) \right]_{+ (1)}^{[-\infty]}  \right. \\
	& {\hskip -0.1in} \left. +\sum_{j=1}^{3} C_j^{(2)} \left[ \theta(-y) \theta(y+1) \tilde{b}_{0j3}^{(2)} (y) \right]_{+ (1)}^{[-\infty]} +\sum_{j=1}^{3} C_j^{(2)} \left[ \theta(-y-1) \tilde{b}_{0j4}^{(2)} (y) \right]_{+ (1)}^{[-\infty]}  \right\}  \,,
	\end{split}
	\end{align}
	where $L_{\delta,0}(y) \equiv \delta(y-1)$, $L_{\delta,1}(y) \equiv L_p \delta(y-1) +\gamma_E \left[ \frac{1}{(1-y)^2} \delta^+\left( \frac{1}{1-y} \right) +\frac{1}{(y-1)^2} \delta^+\left( \frac{1}{y-1} \right) \right] +\left[ \theta(y-1) \left( \frac{1}{1-y} \right) \right]_{\oplus (1,\infty)}^{[2]} +\left[ \theta(1-y) \left( \frac{1}{y-1} \right) \right]_{+ (1,-\infty)}^{[0]}$, $L_{\delta,2}(y) \equiv \left( L_p^2 +\frac{\pi^2}{3} \right) \delta(y-1) +\left( 2 \gamma_E L_p +2 \gamma_E^2 -\frac{\pi^2}{6} \right) \left[ \frac{1}{(1-y)^2} \delta^+\left( \frac{1}{1-y} \right) +\frac{1}{(y-1)^2} \delta^+\left( \frac{1}{y-1} \right) \right] +\left[ \theta(y-1) \left( 2L_p \frac{1}{1-y} +4\frac{\ln(y-1)}{y-1} \right) \right]_{\oplus (1,\infty)}^{[2]} +\left[ \theta(1-y) \left( 2L_p \frac{1}{y-1} +4\frac{\ln(1-y)}{1-y} \right) \right]_{+ (1,-\infty)}^{[0]}$, and $L_p \equiv \ln(\mu^2/4p_z^2)$.
	The complete results are
	\begin{align}\label{eq:qK1K2}
	& \tilde{a}_{011}^{(1)}(y) = \,\frac{2y+1}{2(y-1)} -\frac{(y^2+1)\text{H}(1;1/y)}{y-1} \,,
	{\hskip 0.3in} \tilde{a}_{112}^{(1)}(y) = \,\frac{y^2+1}{y-1} \,,
	{\hskip 0.3in} \tilde{a}_{012}^{(1)}(y) = \,\frac{2y^2+2y-3}{2(y-1)} \nonumber \\
	& \ -\frac{y^2+1}{y-1} \Big[ \text{H}(0;y) -\text{H}(1;y) \Big] \,,
	{\hskip 0.3in} \tilde{a}_{013}^{(1)}(y) = -\frac{2y+1}{2(y-1)} -\frac{y^2+1}{y-1} \Big[ \text{H}(-1;-y) -\text{H}(0;-y) \Big] \,,
	{\hskip 0.3in} \tilde{a}_{111}^{(2)}(y) = \,\frac{3(4y-1)}{8(y-1)} \nonumber \\
	& \ -(y-1)\text{H}(1;1/y) -\frac{(y^2+3)\text{H}(0,1;1/y)}{2(y-1)} -\frac{2(y^2+1)\text{H}(1,1;1/y)}{y-1} \,,
	{\hskip 0.3in} \tilde{a}_{121}^{(2)}(y) = \,\frac{11(2y+1)}{24(y-1)} -\frac{11(y^2+1)\text{H}(1;1/y)}{12(y-1)} \,, \nonumber \\
	& \tilde{a}_{131}^{(2)}(y) = -\frac{2y+1}{6(y-1)} +\frac{(y^2+1)\text{H}(1;1/y)}{3(y-1)} \,,
	{\hskip 0.3in} \tilde{a}_{011}^{(2)}(y) = \,\frac{3(8y+3)}{16(y-1)} -2(y+1)\text{H}(-1;1/y) +\frac{3(4y-1)\text{H}(0;1/y)}{4(y-1)} \nonumber \\
	& \ +\frac{(2y^2-4y+11)\text{H}(1;1/y)}{4(y-1)} -(y-1)\text{H}(0,-1;1/y) -\frac{(3y^2-8y+12)\text{H}(0,1;1/y)}{2(y-1)} -2(y-1)\text{H}(1,0;1/y) \nonumber \\
	& \ -\frac{4(y^2-y+1)\text{H}(1,1;1/y)}{y-1} -\frac{(y^2+7)\text{H}(0,0,1;1/y)}{2(y-1)} -\frac{(y^2+3)\text{H}(0,1,0;1/y)}{y-1} -\frac{2(2y^2+3)\text{H}(0,1,1;1/y)}{y-1} \nonumber \\
	& \ -\frac{y^2+1}{y-1} \Big[ \text{H}(0,0,-1;1/y) +2\text{H}(1,0,-1;1/y) +5\text{H}(1,0,1;1/y) +4\text{H}(1,1,0;1/y) +6\text{H}(1,1,1;1/y) \Big] \,, \nonumber \\
	& \tilde{a}_{021}^{(2)}(y) = \,\frac{604y-223}{144(y-1)} +(y+1)\text{H}(-1;1/y) +\frac{11(2y+1)\text{H}(0;1/y)}{12(y-1)} -\frac{(220y^2-240y+121)\text{H}(1;1/y)}{36(y-1)} \nonumber \\
	& \ +\frac{(y-1)\text{H}(0,-1;1/y)}{2} -\frac{(17y^2+5)\text{H}(0,1;1/y)}{12(y-1)} -\frac{y^2+1}{y-1} \bigg[ \frac{11\text{H}(1,0;1/y)}{6} +\frac{11\text{H}(1,1;1/y)}{6} -\frac{\text{H}(0,0,-1;1/y)}{2}  \nonumber \\
	& \ +\frac{\text{H}(0,0,1;1/y)}{2} -\text{H}(1,0,-1;1/y) \bigg] \,,
	{\hskip 0.3in} \tilde{a}_{031}^{(2)}(y) = -\frac{11(4y-1)}{36(y-1)} -\frac{(2y+1)\text{H}(0;1/y)}{3(y-1)} +\frac{(14y^2-12y+5)\text{H}(1;1/y)}{9(y-1)} \nonumber \\
	& \ +\frac{y^2+1}{y-1} \bigg[ \frac{\text{H}(0,1;1/y)}{3} +\frac{2\text{H}(1,0;1/y)}{3} +\frac{2\text{H}(1,1;1/y)}{3} \bigg] \,,
	{\hskip 0.3in} \tilde{a}_{212}^{(2)}(y) = \,\frac{y-1}{2} +\frac{(3y^2+1)\text{H}(0;y)}{4(y-1)} +\frac{(y^2+1)\text{H}(1;y)}{y-1} \,, \nonumber \\
	& \tilde{a}_{222}^{(2)}(y) = \,\frac{11(y^2+1)}{24(y-1)} \,,
	{\hskip 0.3in} \tilde{a}_{232}^{(2)}(y) = -\frac{y^2+1}{6(y-1)} \,,
	{\hskip 0.3in} \tilde{a}_{112}^{(2)}(y) = -\frac{4y^2-20y+25}{8(y-1)} +\frac{(3y^2+4y-6)\text{H}(0;y)}{2(y-1)} \nonumber \\
	& \ +\frac{(3y^2-2y+3)\text{H}(1;y)}{y-1} -\frac{2(2y^2+1)\text{H}(0,0;y)}{y-1} +\frac{(y+1)\text{H}(0,1;y)}{2} -\frac{y^2+1}{y-1} \Big[ \text{H}(1,0;y) -4\text{H}(1,1;y) \Big] \,, \nonumber \\
	& \tilde{a}_{122}^{(2)}(y) = -\frac{(y^2+1)\pi^2}{12(y-1)} +\frac{440y^2-414y+275}{72(y-1)} -\frac{(y+1)\text{H}(0;y)}{2} +\frac{y^2+1}{y-1} \bigg[ \frac{11\text{H}(1;y)}{12} +\frac{\text{H}(0,0;y)}{2} \bigg] \,, \nonumber \\
	& \tilde{a}_{132}^{(2)}(y) = -\frac{28y^2-18y+13}{18(y-1)} -\frac{(y^2+1)\text{H}(1;y)}{3(y-1)} \,,
	{\hskip 0.3in} \tilde{a}_{012}^{(2)}(y) = -\frac{4(y^2+1)\zeta_3}{y-1} -\frac{(3y^3-y^2-2y+3)\pi^2}{6(y-1)} \nonumber \\
	& \ -\frac{(y^2+3)\pi^2\text{H}(0;y)}{12(y-1)} +\frac{(y^2+1)\pi^2\text{H}(1;y)}{6(y-1)} -\frac{12y^2-64y+69}{16(y-1)} -2(y+1)\text{H}(-1;y) +\frac{(10y^2-17y+1)\text{H}(0;y)}{4(y-1)} \nonumber \\
	& \ +\frac{(7y^2-y-7)\text{H}(1;y)}{4(y-1)} +2(y+1)\text{H}(-1,0;y) +(y-1)\text{H}(0,-1;y) -\frac{(4y^3+32y^2+2y-25)\text{H}(0,0;y)}{4(y-1)} \nonumber \\
	& \ -\frac{(2y^3-6y^2+6y+3)\text{H}(0,1;y)}{2(y-1)} -\frac{(2y^3+5y^2-4y+1)\text{H}(1,0;y)}{2(y-1)} -\frac{(y^3-7y^2+5y-7)\text{H}(1,1;y)}{y-1} \nonumber \\
	& \ +\frac{7(5y^2+3)\text{H}(0,0,0;y)}{4(y-1)} -\frac{(y+1)\text{H}(0,0,1;y)}{2} +\frac{(y^2-3)\text{H}(0,1,1;y)}{2(y-1)} -\frac{y^2+1}{y-1} \bigg[ 2\text{H}(0,-1,0;y) -\text{H}(0,0,-1;y) \nonumber \\
	& \ +\text{H}(0,1,0;y) -2\text{H}(1,0,-1;y) +\frac{3\text{H}(1,0,0;y)}{2} +\text{H}(1,0,1;y) +\text{H}(1,1,0;y) -6\text{H}(1,1,1;y) \bigg] \,, \nonumber \\
	& \tilde{a}_{022}^{(2)}(y) = -\frac{3(y^2+1)\zeta_3}{4(y-1)} -\frac{(y^2-12y-8)\pi^2}{72(y-1)} +\frac{y^2+1}{y-1} \bigg[ \frac{\pi^2\text{H}(0;y)}{12} -\frac{\pi^2\text{H}(1;y)}{3} \bigg] +\frac{5044y^2-4476y+2089}{432(y-1)} \nonumber \\
	& \ +(y+1)\text{H}(-1;y) -\frac{(161y^2-183y+89)\text{H}(0;y)}{36(y-1)} +\frac{(193y^2-204y+112)\text{H}(1;y)}{36(y-1)} -(y+1)\text{H}(-1,0;y) \nonumber \\
	& \ -\frac{(y-1)\text{H}(0,-1;y)}{2} +\frac{(67y^2-24y-35)\text{H}(0,0;y)}{24(y-1)} +\frac{(2y^2+6y+17)\text{H}(0,1;y)}{12(y-1)} +\frac{(y+3)\text{H}(1,0;y)}{4} \nonumber \\
	& \ +\frac{(19y^2+6y+19)\text{H}(1,1;y)}{12(y-1)} +\frac{y^2+1}{y-1} \bigg[ \text{H}(0,-1,0;y) -\frac{\text{H}(0,0,-1;y)}{2} -\frac{7\text{H}(0,0,0;y)}{4} +\frac{\text{H}(0,0,1;y)}{2} -\frac{\text{H}(0,1,0;y)}{2}  \nonumber \\
	& \ -\text{H}(1,0,-1;y) +\text{H}(1,0,0;y) +\frac{\text{H}(1,0,1;y)}{2} -\frac{3\text{H}(1,1,0;y)}{2} \bigg] \,,
	{\hskip 0.3in} \tilde{a}_{032}^{(2)}(y) = -\frac{(y^2+1)\pi^2}{18(y-1)} -\frac{302y^2-168y+71}{108(y-1)} \nonumber \\
	& \ +\frac{(11y^2-12y+11)\text{H}(0;y)}{18(y-1)} -\frac{(14y^2-12y+5)\text{H}(1;y)}{9(y-1)} +\frac{y^2+1}{y-1} \bigg[ \frac{\text{H}(0,0;y)}{6} -\frac{\text{H}(0,1;y)}{3} -\frac{2\text{H}(1,1;y)}{3} \bigg] \,, \nonumber \\
	& \tilde{a}_{113}^{(2)}(y) = \,\frac{(y+1)\pi^2}{12} +\frac{16y^2-12y-13}{8(y-1)} -(y-1)\text{H}(-1;-y) +\frac{2(y^2+1)\text{H}(-1,-1;-y)}{y-1} -\frac{(3y^2+1)\text{H}(0,-1;-y)}{2(y-1)} \nonumber \\
	& \ +\frac{(y+1)\text{H}(0,0;-y)}{2} \,,
	{\hskip 0.3in} \tilde{a}_{123}^{(2)}(y) = -\frac{(y^2+1)\pi^2}{12(y-1)} -\frac{24y^2+22y-13}{24(y-1)} +\frac{(17y^2-12y+17)\text{H}(0;-y)}{12(y-1)} \nonumber \\
	& \ -\frac{(y^2+1)}{y-1} \bigg[ \frac{11\text{H}(-1;-y)}{12} +\text{H}(-1,0;-y) -\frac{\text{H}(0,0;-y)}{2} \bigg] \,,
	{\hskip 0.3in} \tilde{a}_{133}^{(2)}(y) = \,\frac{2y+1}{6(y-1)} +\frac{y^2+1}{y-1} \bigg[ \frac{\text{H}(-1;-y)}{3} -\frac{\text{H}(0;-y)}{3} \bigg] \,, \nonumber \\
	& \tilde{a}_{013}^{(2)}(y) = -\frac{(y^2-3)\zeta_3}{2(y-1)} +\frac{(4y^2-2y-5)\pi^2}{12(y-1)} -\frac{(3y^2+1)\pi^2\text{H}(0;-y)}{12(y-1)} +\frac{44y^2-24y-53}{16(y-1)} +\frac{(2y^2-4y+11)\text{H}(-1;-y)}{4(y-1)} \nonumber \\
	& \ -\frac{(7y^2-14y+3)\text{H}(0;-y)}{4(y-1)} +2(y+1)\text{H}(1;-y) +\frac{4(y^2-y+1)\text{H}(-1,-1;-y)}{y-1} +(3y+1)\text{H}(-1,0;-y) \nonumber \\
	& \ -\frac{(5y^2-4)\text{H}(0,-1;-y)}{2(y-1)} -\frac{(5y^2-4y+2)\text{H}(0,0;-y)}{2(y-1)} -(y-1)\text{H}(0,1;-y) -(y+1)\text{H}(1,0;-y) \nonumber \\
	& \ +\frac{2y^2\text{H}(0,-1,-1;-y)}{y-1} +\frac{(5y^2+3)\text{H}(0,0,-1;-y)}{2(y-1)} -\frac{(y^2-2)\text{H}(0,0,0;-y)}{y-1} -\frac{y^2+1}{y-1} \Big[ 6\text{H}(-1,-1,-1;-y) \nonumber \\
	& \ -\text{H}(-1,0,-1;-y) -2\text{H}(-1,0,1;-y) +\text{H}(0,-1,0;-y) +\text{H}(0,0,1;-y) -\text{H}(0,1,0;-y) \Big] \,, \nonumber \\
	& \tilde{a}_{023}^{(2)}(y) = -\frac{3(y^2+1)\zeta_3}{4(y-1)} -\frac{(2y^2-3y-4)\pi^2}{36(y-1)} +\frac{y^2+1}{y-1} \bigg[ \frac{\pi^2\text{H}(-1;-y)}{12} +\frac{\pi^2\text{H}(0;-y)}{12} \bigg] -\frac{198y^2+604y-421}{144(y-1)} \nonumber \\
	& \ -\frac{(220y^2-240y+121)\text{H}(-1;-y)}{36(y-1)} +\frac{(485y^2-330y+209)\text{H}(0;-y)}{72(y-1)} -(y+1)\text{H}(1;-y) -\frac{(5y^2-2y+1)\text{H}(-1,0;-y)}{2(y-1)} \nonumber \\
	& \ -\frac{(5y^2+17)\text{H}(0,-1;-y)}{12(y-1)} +\frac{(y^2+12y-23)\text{H}(0,0;-y)}{12(y-1)} +\frac{(y-1)\text{H}(0,1;-y)}{2} +\frac{(y+1)\text{H}(1,0;-y)}{2} \nonumber \\
	& \ +\frac{y^2+1}{y-1} \bigg[ \frac{11\text{H}(-1,-1;-y)}{6} +\text{H}(-1,-1,0;-y) +\frac{3\text{H}(-1,0,0;-y)}{2} -\text{H}(-1,0,1;-y) +\text{H}(0,-1,0;-y) \nonumber \\
	& \ -\frac{\text{H}(0,0,-1;-y)}{2} -\frac{7\text{H}(0,0,0;-y)}{4} +\frac{\text{H}(0,0,1;-y)}{2} -\frac{\text{H}(0,1,0;-y)}{2} \bigg] \,,
	{\hskip 0.3in} \tilde{a}_{033}^{(2)}(y) = -\frac{(y^2+1)\pi^2}{18(y-1)} +\frac{11(4y-1)}{36(y-1)} \nonumber \\
	& \ +\frac{(14y^2-12y+5)\text{H}(-1;-y)}{9(y-1)} -\frac{2(7y^2-3y+4)\text{H}(0;-y)}{9(y-1)} -\frac{y^2+1}{y-1} \bigg[ \frac{2\text{H}(-1,-1;-y)}{3} -\frac{\text{H}(0,-1;-y)}{3} -\frac{\text{H}(0,0;-y)}{3} \bigg] \,, \nonumber \\
	& \tilde{b}_{012}^{(1)}(y) = \,2(y-1) \,,
	{\hskip 0.3in} \tilde{b}_{011}^{(2)}(y) = \,\frac{3(4y-5)}{4(y-1)} -(y-1) \Big[ 2\text{H}(1;1/y) +\text{H}(0,1;1/y) +2\text{H}(1,1;1/y) \Big] \,,  \nonumber \\
	& \tilde{b}_{021}^{(2)}(y) = \,\frac{11(2y-3)}{12(y-1)} -\frac{11(y-1)\text{H}(1;1/y)}{6} \,,
	{\hskip 0.3in} \tilde{b}_{031}^{(2)}(y) = -\frac{2y-3}{3(y-1)} +\frac{2(y-1)\text{H}(1;1/y)}{3} \,,
	{\hskip 0.3in} \tilde{b}_{112}^{(2)}(y) = \,2(y-1) \nonumber \\
	& \ +(y-1) \Big[ \text{H}(0;y) +2\text{H}(1;y) \Big] \,,
	{\hskip 0.3in} \tilde{b}_{122}^{(2)}(y) = \,\frac{11(y-1)}{6} \,,
	{\hskip 0.3in} \tilde{b}_{132}^{(2)}(y) = -\frac{2(y-1)}{3} \,,
	{\hskip 0.3in} \tilde{b}_{012}^{(2)}(y) = -\frac{(3y^2-2y)\pi^2}{6} \nonumber \\
	& \ -\frac{3(8y^2-20y+11)}{4(y-1)} +3(y+3)\text{H}(0;y) +5(y-1)\text{H}(1;y) -(y^2+6y-4)\text{H}(0,0;y) -(y^2-y+1)\text{H}(0,1;y) \nonumber \\
	& \ -(y^2+4y-3)\text{H}(1,0;y) -(y^2-4y+3)\text{H}(1,1;y) \,,
	{\hskip 0.3in} \tilde{b}_{022}^{(2)}(y) = \,\frac{\pi^2}{6} +\frac{410y^2-754y+377}{36(y-1)} -2(y+1)\text{H}(0;y) \nonumber \\
	& \ +\frac{5(y-1)\text{H}(1;y)}{6} +(2y-1)\text{H}(0,0;y) +y\text{H}(1,0;y) \,,
	{\hskip 0.3in} \tilde{b}_{032}^{(2)}(y) = -\frac{22y^2-38y+19}{9(y-1)} -\frac{2(y-1)\text{H}(1;y)}{3} \,, \nonumber \\
	& \tilde{b}_{013}^{(2)}(y) = \,\frac{(3y-1)\pi^2}{6} +\frac{3(8y^2-4y-3)}{4(y-1)} +2\text{H}(0;-y) +2(3y-1)\text{H}(-1,0;-y) -(5y-1)\text{H}(0,0;-y) \nonumber \\
	& \ -(y-1) \Big[ 2\text{H}(-1;-y) -2\text{H}(-1,-1;-y) +\text{H}(0,-1;-y) \Big] \,,
	{\hskip 0.3in} \tilde{b}_{023}^{(2)}(y) = -\frac{(2y-1)\pi^2}{6} -\frac{36y^2+22y-69}{12(y-1)} \nonumber \\
	& \ -\frac{11(y-1)\text{H}(-1;-y)}{6} +\frac{(17y-23)\text{H}(0;-y)}{6} -2(2y-1)\text{H}(-1,0;-y) +(3y-1)\text{H}(0,0;-y) \,, \nonumber \\
	& \tilde{b}_{033}^{(2)}(y) = \,\frac{2y-3}{3(y-1)} +(y-1) \bigg[ \frac{2\text{H}(-1;-y)}{3} -\frac{2\text{H}(0;-y)}{3} \bigg] \,,
	\end{align}
	and $\tilde{a}_{ij4}^{(n)}(y) = -\tilde{a}_{ij1}^{(n)}(y)$, $\tilde{b}_{ij4}^{(n)}(y) = -\tilde{b}_{ij1}^{(n)}(y)$.
	Additional harmonic polylogarithms here are as follows ($0 < z < 1$):
	\begin{align}\label{eq:HPL1}
	& \text{H}(-1,-1;z) = \,\frac{1}{2}\ln^2(1+z) \,,
	{\hskip 0.3in} \text{H}(0,-1;z) = -\text{Li}_2\left(-z\right) \,,
	{\hskip 0.3in} \text{H}(-1,-1,-1;z) = \,\frac{1}{6}\ln^3(1+z) \,, \nonumber \\
	& \text{H}(-1,0,-1;z) = -2\zeta_3 +\frac{\pi^2}{3} \ln(1+z) -\frac{1}{3}\ln^3(1+z) +\ln(1+z) \text{H}(-1,0;z) +2\text{Li}_3\left(\frac{1}{1+z}\right) \,, \nonumber \\
	& \text{H}(-1,0,1;z) = \,\zeta_3 -2\text{Li}_3\left(1/2\right) -\text{Li}_2\left(1/2\right) \Big[ \ln(1-z) +\ln(1+z) \Big] -\ln(2)\ln(1-z)\ln(1+z) +\frac{1}{2}\ln(1-z)\ln^2(1+z) \nonumber \\
	& \ -\ln(1-z) \text{H}(-1,0;z) +\text{Li}_3\left(\frac{1-z}{2}\right) -\text{Li}_3\left(1-z\right) -\text{Li}_3\left(\frac{1}{1+z}\right) +\text{Li}_3\left(\frac{1-z}{1+z}\right) +\text{Li}_3\left(\frac{1+z}{2}\right) \,,  \nonumber \\
	& \text{H}(0,-1,-1;z) = \,\zeta_3 -\frac{\pi^2}{6} \ln(1+z) -\frac{1}{2}\ln(z)\ln^2(1+z) +\frac{1}{6}\ln^3(1+z) -\ln(1+z) \text{Li}_2\left(-z\right) -\text{Li}_3\left(\frac{1}{1+z}\right) \,, \nonumber \\
	& \text{H}(0,0,-1;z) = -\text{Li}_3\left(-z\right) \,,
	{\hskip 0.3in} \text{H}(0,0,1;z) = \,\text{Li}_3\left(z\right) \,,
	{\hskip 0.3in} \text{H}(1,0,-1;z) = -\frac{21}{4}\zeta_3 +\frac{\pi^2}{3}\ln(2) +6\text{Li}_3\left(1/2\right) \nonumber \\
	& \ +2\ln(2) \text{Li}_2\left(1/2\right) +\ln(1-z) \text{H}(-1,0;z) +\ln(1+z) \text{H}(1,0;z) +\text{H}(-1,0,1;z) \,. 
	\end{align}
	
	Complete analytical expressions of $\tilde{a}_{ijk}^{(n)}$ and $\tilde{b}_{ijk}^{(n)}$ and a code to calculate $i \tilde{K}^\sigma \left( y, 1/p_z^2, \mu^2 \right)$ when $y \in (-\infty,0)\cup(0,1)\cup(1,\infty)$ are available to download from an ancillary file.

\end{widetext}

\providecommand{\href}[2]{#2}\begingroup\raggedright\endgroup

\end{document}